\documentclass[journal,english]{IEEEtran}

\usepackage[style=ieee, doi=false, isbn=false, url=false]{biblatex}
\usepackage{amsmath,amsfonts,algorithmic,algorithm}
\usepackage{array}
\usepackage{textcomp}
\usepackage{stfloats}
\usepackage{url}
\usepackage{hyperref}
\usepackage{verbatim}
\usepackage{graphicx}
\usepackage{amssymb}
\usepackage{mathtools}
\usepackage{amsthm}
\usepackage{xcolor}
\usepackage{physics}
\usepackage{stmaryrd}
\usepackage{bm}
\usepackage{epstopdf}
\usepackage{subfigure}
\usepackage{bbm}
\usepackage[normalem]{ulem}
\usepackage{enumitem}
\usepackage{babel,csquotes}

\allowdisplaybreaks

\DeclareMathOperator*{\argmin}{\arg\,min}
\DeclareMathOperator*{\argmax}{\arg\,max}

\newcommand{\ignore}[1]{}

\newcommand\herm{{\mathsf{H}}}

\def\BibTeX{{\rm B\kern-.05em{\sc i\kern-.025em b}\kern-.08em
    T\kern-.1667em\lower.7ex\hbox{E}\kern-.125emX}}

\begin{document}

\title{Channel State Information-Free Angle Spoofing for Differential Localization Privacy}

\author{
    \IEEEauthorblockN{Lei Lyu and Urbashi Mitra}\\
    \IEEEauthorblockA{University of Southern California, Los Angeles, USA}
    \thanks{This work has been funded in part by one or more of the following: ARO W911NF1910269, ARO W911NF2410094,  NSF CIF-2311653, NSF RINGS CIF-2148313,  ONR N00014-22-1-2363, NSF DBI-2412522, and is also supported in part by funds from federal agency and industry partners as specified in the Resilient \& Intelligent NextG Systems (RINGS) program.}
}

\maketitle

\begin{abstract}
This paper proposes a channel state information (CSI)-free angle spoofing strategy that remains effective even when the eavesdropper (Eve) uses differential statistics to estimate the location of the transmitter (Alice). Specifically, by designing a precoding matrix, Eve incorrectly estimates the angle differences of departure (ADoDs), thereby spoofing Alice's location. Meanwhile, the trusted user (Bob) can still infer Alice's true location with a modest amount of securely shared side information. Given the channel parameters estimated by Eve from the spoofed channel, the resulting localization bias introduced to the eavesdroppers under the spoofed geometric is analyzed. Eve's bias can be driven to infinity asymptotically in a key parameter of the precoder. Simulation results demonstrate that the proposed method introduces significant localization errors for Eve. When combined with a prior CSI-free spoofing strategy, further degraded performance is achieved by Eve.
\end{abstract}


\section{Introduction}
Due to the large bandwidth and high resolution, millimeter wave (mmWave) signals~\cite{rappaport2013millimeter} are widely employed to provide high-accuracy localization. Recent channel estimation algorithms~\cite{fascista2021downlink, li2022joint, pourafzal2025flex} accurately recover {multidimensional multipath channel parameters} including {times of arrivals (ToAs)}, angles of departure (AoDs) and angles of arrival (AoAs), which can be exploited for localization. While these methods enable accurate positioning, they can also be used by eavesdroppers, allowing unauthorized inference of the transmitter's location.

% Protecting location privacy is a fundamental requirement for wireless localization applications. 
%To combat this issue,
%physical-layer-based privacy preserving techniques have been investigated. 
Artificial-noise-based methods enhance physical-layer security by degrading the eavesdropper's channel~\cite{goel2008guaranteeing, tomasin2022beamforming}.
% \um{we should look for more recent methods}
Beamforming techniques have also been proposed to hide critical location information from {Eve~\cite{checa2020location, ayyalasomayajula2023users}{, or limit Eve's Cram\'er--Rao Bound}~\cite{zhang2025privacy}}.
% , thereby reducing the risk of position leakage
% In~\cite{zhang2025privacy}, beamformers are designed by solving an optimization problem that minimizes the legitimate localization Cramer-Rao Bound (CRB), while constraining the eavesdropper's CRB. 
A power allocation strategy is proposed in~\cite{zhu2025securing} to limit position exposure to Eve.
%where the transmitted power from an agent to anchors or other agents is optimized to ensure the localization performance while limiting the position exposure {to Eve}. 
However, these methods rely on known CSI, which is generally unavailable since eavesdroppers do not disclose their locations.  
% \um{what is wrong with these methods?}

%Location spoofing has emerged as an effective physical-layer strategy for protecting location information. Existing methods, also typically require some {CSI knowledge which is generally unavailable since eavesdroppers do not disclose their locations. 
Recent location spoofing methods also require (some) CSI.
Angle information of the channel between Eve and Alice is required  %to spoof targeted AoAs or AoDs 
in~\cite{srinivasan2024aoa,maity2026pilot}. In~\cite{arun2024dolos}, the direct and reflected channel components are needed to selectively shift the ToA of the direct path.
% For example, the method in~\cite{srinivasan2024aoa} requires to know the AoAs of Eve's channel, which are generally unavailable since eavesdroppers do not disclose their locations. 
To address this limitation, a CSI-free delay-angle information spoofing (DAIS) method was proposed in~\cite{li2024channel, li2024optimized, li2025delay}, where a precoding matrix is designed to shift Eve's ToA and AoD estimates. However, DAIS is less effective in the presence of  \emph{differential-signaling-based} localization 
% \um{officially introduced in the ICC paper I think you will also want to cite the ISIT paper since these two are published into addition to the arXiV work}
% \um{are there other differntial methods?  Doesn't Badria from UCSD have one?} 
{techniques which are common in the literature.} {Specifically, time differences of arrival (TDoAs) are used %are used instead of ToAs 
in~\cite{palacios2022low,arun2023xrloc, shamsian2023joint}, and~\cite{chen2023learning} uses TDoAs, ADoDs and angle differences of arrival (ADoAs) for distance estimation. 
%As shown therein, DAIS-induced spoofing can {be insufficient sometimes} when using 
%TDoAs and ADoDs.} \um {in~\cite{palacios2022low,arun2023xrloc, shamsian2023joint}, and~\cite{chen2023learning} show that DAIS does not work??}

Therefore, in this paper, we design a precoding-based differential-signaling angle spoofing (DSAS) strategy that can mislead Eve into learning incorrect {ADoDs}. The main contributions of this paper are as follows:
\begin{enumerate}[nosep]
    \item A precoding matrix is designed to spoof Eve's estimates of {AoDs as well as the relative AoDs between non-line-of-sight (NLoS) paths and the line-of-sight (LoS) path.}
    \item {The effectiveness of DAIS~\cite{li2024channel, li2024optimized, li2025delay} against differential localization~\cite{chen2023learning} is analyzed {and simulated}.}
    \item {In the absence of noise, {approximate spoofed ADoDs for Eve and Eve's localization error are derived}. It is shown that the bias for Eve's estimates can be driven to be large through the careful precoding parameter choice.}
    % \um{I think we need to use different language, all estimates are approximate so what does an approximation of an estimate mean?}
    \item Simulation results demonstrate the effectiveness of DSAS, {and its performance improvement when combined with} DAIS under differential localization. 
\end{enumerate}
The following notational conventions are adopted: lower case $x$ denotes scalars, {a variant of $\mathrm{x}$ denotes random scalars}, boldface lower case $\bm{x}$ denotes vectors, a variant of the boldface $\bm{\mathrm{x}}$ denotes random vectors and boldface upper case $\bm{X}$ denotes matrices. {The $(t, k)$-th entry of the matrix $\bm{X}$ is denoted by $X_{t, k}$.}
% The vector from $t_1$-th to $t_2$-th entry of $\bm{x}$ is denoted as $\bm{x}_{t_1:t_2}$, where $x_t \doteq \bm{x}_{t:t}$. The matrix from $t_1$-th to $t_2$-th row and $k_1$-th to $k_2$-th column of $\bm{X}$ is denoted $\bm{X}_{t_1:t_2, k_1:k_2}$, where $X_{t, k} \doteq \bm{X}_{t:t, k:k}$. 
We denote $x^\star$ and $\bm{x}^\star$ as the {\bf ground-truth} values of scalar and vector parameters. The quantities $\hat{x}\left(\cdot\right)$ and $\hat{\bm{x}}\left(\cdot\right)$ denote estimates of scalar and vector parameters, respectively. The operators $\left(\cdot\right)^{\top}$, $\overline{\left(\cdot\right)}$, $\left(\cdot\right)^{\herm}$, and $\left(\cdot\right)^\dagger$ denote transpose, conjugate, conjugate transpose, and pseudoinverse, respectively. The Hadamard product is given by $\odot$.
% The integer-valued interval is denoted by $\llbracket a, b\rrbracket =\left\{c \in \mathbb{Z}:a\leq c \leq b\right\}$.
% \um{put in some sentences explaining your notation}

\section{Preliminaries}
\vspace*{-0.02in}
\subsection{Wireless Channel Models}
\vspace*{-0.05in}

% \um{Lei, you need to be clear about what the dimensions of key vectors and matrices are}
We consider a multiple-input multiple-output (MIMO)  channel with orthogonal frequency-division multiplexing (OFDM) signaling. The frequency-domain channel over the $n$-th subcarrier is given by
\begin{align}
    \bm{H}^{(n)}&=\sum\limits_{k=0}^K\gamma_k^\star e^{\frac{-j2\pi n\tau_k^\star}{NT_s}}\bm{A}_{N_R, N_T}\left(\theta_{Rx, k}^\star, \theta_{Tx, k}^\star\right),
\end{align}
for $n \in \left \{ 1, \cdots, N\right \}$, where $K+1$, $N_R$ and $N_T$ represent the numbers of propagation paths, receivers, and transmitters, respectively. The parameters $\gamma_k^\star$, $\tau_k^\star$, $\theta_{Rx, k}^\star$ and $\theta_{Tx, k}^\star$ represent the ground-truth amplitude, ToA, AoA and AoD for the $k$-th path, where {$\tau_0^\star \leq \tau_1^\star\leq\cdots\leq \tau_K^\star${,} and the line-of-sight (LoS) path is indexed by $0$.} For $\vartheta, \varphi\in \left(-\frac{\pi}{2}, \frac{\pi}{2}\right)$, the steering matrix {for uniform linear antenna arrays at both the transmitter and receiver} is defined as
\begin{equation}
    \bm{A}_{N_R, N_T}\left(\vartheta, \varphi\right) \doteq \bm{a}_{N_R}\left(\vartheta\right)\bm{a}_{N_T}^\herm\left(\varphi\right)\in \mathbb{C}^{N_R\times N_t},
    \label{eq:A_mat}
\end{equation}
where the exponential steering vector takes the form
{
\begin{equation}
    \bm{a}_{N_I}\left(\vartheta\right) \doteq \frac{1}{N_I}\left[1, e^{-\mathrm{j}\frac{d}{\lambda_c}2\pi\sin\vartheta}, \cdots, e^{-\mathrm{j}\frac{d}{\lambda_c}2\pi\left(N_I-1\right)\sin\vartheta}\right]^\top,
\end{equation}
for $N_I\in \left \{N_R, N_T\right \}$ and $\vartheta\in \left(-\frac{\pi}{2}, \frac{\pi}{2}\right)$. The distance between antennas, $d$, is set to half the wavelength $\lambda_c$, i.e. $d=\frac{\lambda_c}{2}$.}

% \um{the subscript of $r_1 n r_2$ looks odd, I would do $n r_1 r_2$}

% We assume $N \times R_1\times  R_2$ independent scalar observations~\cite{pourafzal2025flex}, {
% \begin{align}
%     \mathrm{y}_{n, r_1, r_2} &= \bm{w}_{r_1}^\herm \bm{H}^{(n)}\bm{f}_{r_2}s_{n, r_1, r_2}+\mathrm{e}_{n, r_1, r_2},
% \end{align}
% where $\mathrm{e}_{n, r_1, r_2}\sim \mathcal{CN}\left( 0, \sigma_e^2\right)$ is the noise and $s_{n, r_1, r_2}\in\mathbb{C}$ is the transmitted symbol.} The vectors $\bm{w}_{r_1}\in \mathbb{C}^{N_R}$ and $\bm{f}_{r_2}\in \mathbb{C}^{N_T}$ are a combiner and precoder vectors, respectively, for each $r_1\in \left\{1, \cdots, R_1\right\}$ and $r_2\in \left\{1, \cdots, R_2\right\}$.

{We assume $N$ independent observation blocks, 
\begin{align}
    \bm{\mathrm{Y}}^{(n)} &= \left(\bm{W}^\herm \bm{H}^{(n)}\bm{F}\right)\odot \bm{S}^{(n)}+\bm{\mathrm{E}}^{(n)}\in \mathbb{C}^{R_1\times R_2},
\end{align}
where the noise $\mathrm{E}^{(n)}_{r_1, r_2}\overset{\text{i.i.d}}{\sim} \mathcal{CN}\left( 0, \sigma_e^2\right)$ and $\bm{S}^{(n)}\in\mathbb{C}^{R_1\times R_2}$ is the $n$-th transmitted symbol block. $\bm{W}\in \mathbb{C}^{N_R\times R_1}$ and $\bm{F}\in \mathbb{C}^{N_T \times R_2}$ are the combiner and precoder matrices, respectively.}

% Similar to~\cite{li2025delay}, in the following analyses, set $\bm{w}_{r_1}=\bm{e}_{N_R, r_1}$ and $\bm{f}_{r_2}=\bm{e}_{N_T, r_2}$, where $e_{N,r}$ represent the $r$-th standard basis vector of length $N$, then
% \begin{align}
%     \vb{y}_{n, r_1, r_2} &= \bm{H}^{(n)}\left[r_1, r_2\right]+\vb{w}_{n, r_1, r_2},
% \end{align}
% $r_1\in \left\{1, \cdots, N_R\right\}$ and $r_2\in \left\{1, \cdots, N_T\right\}$.

\subsection{Differential-Signaling Localization Method}
Suppose an authorized user (Alice) is at the location $\bm{p}^\star\in \mathbb{R}^2$. A localizer (Bob/Eve) at $\bm{q}^\star\in \mathbb{R}^2$ is trying to infer the location of Alice, with geometric parameters $\bm{\tau}=\left[\tau_0, \cdots, \tau_K\right]^\top$, $\bm{\theta}_{Rx}=\left[\theta_{Rx, 0}, \cdots, \theta_{Rx, K}\right]^\top$ and $\bm{\theta}_{Tx}=\left[\theta_{Tx, 0}, \cdots, \theta_{Tx, K}\right]^\top$. We define the corresponding \emph{differential parameters} as $\dot{\bm{x}}\doteq \left[\left\vert x_1-x_0\right\vert, \cdots, \left\vert x_K-x_0\right\vert\right]^\top\in \mathbb{R}^K$ for $\bm{x}\in\left\{\bm{\tau}, \bm{\theta}_{Rx}, \bm{\theta}_{Tx}\right\}$. 

According to the differential-signaling localization method~\cite{chen2023learning}, the localizer estimates Alice's location as
\begin{align}
    \bm{p}= \bm{q}^\star-d\left[
        \cos{\theta_{Rx, 0}}, \sin{\theta_{Rx, 0}}\right]^\top\in\mathbb{R}^2,
        \label{eq:ds_loc_original}
\end{align}
where $d$ is {the distance estimate} given by,
\begin{equation}
d=\frac{\left<{v_c}\dot{\bm{\tau}}\odot\sin\left(\dot{\bm{\theta}}_{Rx}+\dot{\bm{\theta}}_{Tx}\right), \bm{f}\left(\dot{\bm{\theta}}_{Rx}, \dot{\bm{\theta}}_{Tx}\right)\right>}{\left\Vert \bm{f}\left(\dot{\bm{\theta}}_{Rx}, \dot{\bm{\theta}}_{Tx}\right)\right\Vert_2^2},
\label{eq:diff_dist_est}
\end{equation}
and the function ${\bm{f}}:\mathbb{R}^l\times\mathbb{R}^l\rightarrow\mathbb{R}^l$ is defined as
\begin{align}
    \bm{f}\left(\bm{\vartheta}, \bm{\varphi}\right)&\doteq\sin\bm{\vartheta}+\sin\bm{\varphi}-\sin\left(\bm{\vartheta}+\bm{\varphi}\right),
\end{align}
where $\sin{\left(\cdot\right)}$ is element-wise for $\bm{\vartheta}, \bm{\varphi}\in \mathbb{R}^l$.
% \um{this sentence does not make sense, if Bob has the groundtruth already, he does not need to estimate Alice's location, what are you really trying to say? With the ground-truth parameters $\bm{\tau}=\bm{\tau}^\star$, $\bm{\theta}_{Rx}=\bm{\theta}_{Rx}^\star$ and $\bm{\theta}_{Tx}=\bm{\theta}_{Tx}^\star$, the localizer can infer the true location of Alice $\bm{p}=\bm{p}^\star$.}

Notice that Equation~\eqref{eq:diff_dist_est} can be expressed as the weighted average over the pairwise distance estimates
\begin{equation}
    d = \sum\limits_{k=1}^K c_k \frac{{v_c}\dot{\tau}_k\sin\left(\dot{\theta}_{Rx, k}+\dot{\theta}_{Tx, k}\right)}{f\left(\dot{\theta}_{Rx, k}, \dot{\theta}_{Tx, k}\right)}\doteq \sum\limits_{k=1}^K c_k d_k,
    \label{eq:diff_dist_est_sum}
\end{equation}
where {$v_c$ denotes the signal propagation speed}, $d_k$ is the distance estimate with the $k$-th LoS-NLoS pair and $c_k =\frac{f^2\left(\dot{\theta}_{Rx, k}, \dot{\theta}_{Tx, k}\right)}{\sum\limits_{k'=1}^K f^2\left(\dot{\theta}_{Rx, k'}, \dot{\theta}_{Tx, k'}\right)}\geq 0$ is the weight.

Using trigonometric equalities, we can further simplify the expression for $d_k$: 
% \um{might be useful to list the exact ones that you employ e.g. sin (a + b) = ...}
\begin{align}
    d_k &\overset{(a)}{=} \frac{{v_c}\dot{\tau}_k\cdot2\sin{\frac{\dot{\theta}_{Rx, k}+\dot{\theta}_{Tx, k}}{2}}\cos{\frac{\dot{\theta}_{Rx, k}+\dot{\theta}_{Tx, k}}{2}}}{2\sin{\frac{\dot{\theta}_{Rx, k}+\dot{\theta}_{Tx, k}}{2}}\left(\cos{\frac{\dot{\theta}_{Rx, k}-\dot{\theta}_{Tx, k}}{2}}-\cos{\frac{\dot{\theta}_{Rx, k}+\dot{\theta}_{Tx, k}}{2}}\right)}\nonumber
    \\
    % &\overset{(b)}{=} \frac{v_c\dot{\tau}_k\cdot\left(\cos{\frac{\dot{\theta}_{Rx, k}}{2}}\cos{\frac{\dot{\theta}_{Tx, k}}{2}}-\sin{\frac{\dot{\theta}_{Rx, k}}{2}}\sin{\frac{\dot{\theta}_{Tx, k}}{2}}\right)}{2\sin{\frac{\dot{\theta}_{Rx, k}}{2}}\sin{\frac{\dot{\theta}_{Tx, k}}{2}}}\nonumber
    % \\
    &\overset{(b)}{=} \frac{{v_c}\dot{\tau}_k}{2}\left(\cot{\frac{\dot{\theta}_{Rx, k}}{2}}\cot{\frac{\dot{\theta}_{Tx, k}}{2}}-1\right),
    \label{eq:diff_dist_est_pair}
\end{align}
where $(a)$ follows by $\sin{\vartheta}+\sin{\varphi}=2\sin{\frac{\vartheta+\varphi}{2}}\cos{\frac{\vartheta-\varphi}{2}}$ and $\sin{\left(2\vartheta\right)}=2\sin{\vartheta}\cos{\vartheta}$, for $\vartheta, \varphi\in \mathbb{R}$. $(b)$ follows by $\cos{\left(\vartheta+\varphi\right)}=\cos{\vartheta}\cos{\varphi}-\sin{\vartheta}\sin{\varphi}$ and $\cos{\vartheta}-\cos{\varphi}=2\sin{\frac{\vartheta+\varphi}{2}}\sin{\frac{\varphi-\vartheta}{2}}$, for $\vartheta, \varphi\in \mathbb{R}$.
% \um{these manipulations above do not occur in [11]?}
%\subsection{Potential {Insufficiency} of DAIS for Eve}

Recall that DAIS~\cite{li2025delay} designs the precoding matrix as 
\begin{equation}
    \bm{\Phi}^{(n)}\left(\Delta_\tau, \Delta_\theta\right)=e^{-\frac{j2\pi n \Delta_\tau}{NT_s}}\mbox{diag}\left\{{\overline{\bm{a}}_{N_T}\left(\Delta_\theta\right)}\right\},
    \label{eq:dais_mat}
\end{equation}
for $\Delta_\tau\in \left(0, NT_s\right]$ and $\Delta_\theta\in \left(0, \pi\right]$. {In the absence of phase wrapping in the ToAs ($\tau_0+\Delta_\tau>NT_s$ or $\tau_K+\Delta_\tau<NT_s$), DAIS can be ineffective as the shift is removed for TDoAs.}
%{The spoofing will be {less effective} under {the above} differential localization, unless phase wrapping occurs in ToAs: $\tau_0+\Delta_\tau<NT_s$ and $\tau_K+\Delta_\tau>NT_s$, which leads to misidentification of the LoS and NLoS paths. The probability of Eve in such a location decreases as the number of paths decreases.}
% ; (ii) $ \min\limits_{k=0, \cdots, K}\theta_{Tx, k}+\Delta_\theta<\frac{\pi}{2}$ and $\max\limits_{k=0, \cdots, K}\theta_{Tx, k}+\Delta_\theta>\frac{\pi}{2}$.
{The impact of shifted AoDs is non-linear and DAIS can still provide spoofing for these terms, but DAIS's overall effectiveness is lowered for differential signaling based localization.}
%{Although the nonlinear AoD shifts result in spoofed ADoD statistics, our numerical results show that DAIS, when restricted to angle spoofing only, is less effective than DSAS on average. Furthermore, angle-spoofing-only DAIS exhibits insufficient performance for certain channel realizations.}
% {Although the nonlinear shift in AoDs results in spoofed ADoD statistics, }from our numerical results in the sequel, we will see that there are channel realizations for which DAIS {is less effective} {under the above differential-signaling localization}, whereas DSAS {always} works. Moreover, the probability that Eve is in such a position increases as the number of paths decreases. {However, DAIS can still introduce significant errors into Eve’s estimates for certain channel realizations -- it does not always fail in differential environments.}
% Whether DAIS or DSAS performs better depends on the specific channel realization.

\section{Differential-Signaling Angle Spoofing}
Our aim is to design a precoding strategy inspired by \cite{li2025delay} that yields a spoofing of Alice's position that is robust to the use of differential-signaling localization \cite{chen2023learning}. Furthermore, the spoofing effect should be achieved without Alice possessing any information about the CSI from herself to the localizer.  

{We observe that $\bm{A}_{N_R, N_T}\left(\vartheta_1, \varphi_1\right)\bm{a}_{N_T}\left(\vartheta_2\right)\bm{a}_{N_T}^\herm\left(\varphi_2\right)=\left[\bm{a}_{N_T}^\herm\left(\varphi_1\right)\bm{a}_{N_T}\left(\vartheta_2\right)\right]\bm{A}_{N_R, N_T}\left(\vartheta_1, \varphi_2\right)$, where the parameter $\varphi_1$ of the steering matrix is spoofed to an arbitrary angle $\varphi_2$.} Motivated by this, we design the following precoding matrix to spoof the AoDs of the channel:
\begin{equation}
    \bm{\Psi}\left(\bm{\phi}_{Tx}^{(n)}\right) = \sum\limits_{m=0}^{M-1}\bm{B}_{N_T}\left(\sin^{-1}{\left(\frac{2m-M}{M}\right)}, \phi_{Tx, m}^{(n)}\right),
\end{equation}
where $M\geq N_T$ is a pre-defined parameter and the matrix  $\bm{B}_{N_T}\left(\vartheta, \varphi\right)$ is defined as $\bm{B}_{N_T}\left(\vartheta, \varphi\right)\doteq \bm{a}_{N_T}\left(\vartheta\right)\bm{a}_{N_T}\left(\varphi\right)^\herm$, $\forall \vartheta, \varphi \in \left(-\frac{\pi}{2}, \frac{\pi}{2}\right)$. The side information which must be shared secretly with Bob is 
 $\bm{\phi}_{Tx}^{(n)}$. We underscore that $\bm{\phi}_{Tx}^{(n)}$ is {\bf not} a function of either $\bm{\theta}_{Rx}^\star$ or $\bm{\theta}_{Tx}^\star$. With precoding, {the $n$-th observation block at Eve is now
\begin{align}
    \widetilde{\bm{\mathrm{Y}}}^{(n)} &= \left(\bm{W}^\herm \bm{H}^{(n)}\bm{\Psi}\left(\bm{\phi}_{Tx}^{(n)}\right)\bm{F}\right)\odot \bm{S}^{(n)}+\bm{\mathrm{E}}^{(n)}.
\end{align}}
We allow the eavesdropper, Eve, to have significant side information, that is: $\bm{W}$, $\bm{F}$ and $\bm{S}^{(n)}$. Then, the channel experienced by Eve is the spoofed channel given by,
% \begin{align}
%     \widetilde{\bm{H}}^{(n)} =&\bm{H}^{(n)}\bm{\Psi}\left(\bm{\phi}_{Tx}^{(n)}\right)\nonumber
%     \\
%     =&\sum\limits_{m=0}^{M-1}\sum\limits_{k=0}^K
%     \gamma_k^\star g_{N_T}\left(\frac{2m-M}{M}- \sin\theta_{Tx, k}^\star\right) \times\nonumber
%     \\
%     &e^{-j2\pi n\frac{\tau_k^\star}{NT_s}}\bm{A}_{N_R, N_T}\left(\theta_{Rx, k}^\star, \phi_{Tx, m}^{( n)}\right),
%     \label{eq:spoof_channel_mul}
% \end{align}
\begin{align}
    &\widetilde{\bm{H}}^{(n)} =\bm{H}^{(n)}\bm{\Psi}\left(\bm{\phi}_{Tx}^{(n)}\right)\nonumber
    \\
    =&\sum\limits_{m=0}^{M-1}\sum\limits_{k=0}^K
    \gamma_k^\star \alpha_{m ,k}e^{\frac{-j2\pi n\tau_k^\star}{NT_s}}\bm{A}_{N_R, N_T}\left(\theta_{Rx, k}^\star, \phi_{Tx, m}^{( n)}\right),
    \label{eq:spoof_channel_mul}
\end{align}
where $\alpha_{m, k}\doteq g_{N_T}\left(\frac{2m-M}{M}- \sin\theta_{Tx, k}^\star\right)$ and the function $g_{N_0}(\omega)\doteq\frac{1-e^{j\pi N_0\omega}}{1-e^{j\pi\omega}}$, $\forall N_0\in \mathbb{Z}\setminus \left\{0\right\}$ and $\omega\in \left(-1, 1\right]$.
% \um{What is the intuition that this all works?}
{We assume Bob knows the precoder structure and can reconstruct $\phi^{(n)}_{Tx, m}$ given the spoofing parameters $\bm{\phi}_{Tx}^{(n)}$. If $M\geq N_T$ and $\bm{\phi}_{Tx}^{(n)}$ is chosen properly, then
$\bm{\Psi}\left(\bm{\phi}_{Tx}^{(n)}\right)$ is invertible, and Bob can recover the correct channel response from $\bm{\Psi}^{-1}\left(\bm{\phi}_{Tx}^{(n)}\right)\widetilde{\bm{H}}^{(n)}$. A sequence of spoofing parameters can be encoded in a codebook that is indexed by a single number shared between Alice and Bob.}

%{For Bob, it suffices to know the mapping $\phi^{(n)}_{Tx, m}$, for $n=1, \cdots, N$ and $m=0, \cdots, M-1$. With this information, Bob can reconstruct the precoder $\bm{\Psi}\left(\bm{\phi}_{Tx}^{(n)}\right)$ and recover the correct channel response as $\bm{\Psi}^{-1}\left(\bm{\phi}_{Tx}^{(n)}\right)\widetilde{\bm{H}}^{(n)}$. Therefore we need to design $\bm{\Psi}\left(\bm{\phi}_{Tx}^{(n)}\right)$ to be invertible, which requires $M\geq N_T$.}

%{Furthermore, the mapping can be encoded into a codebook, so that only a scalar index needs to be shared with Bob. In the simulations, we generate a new mapping for each transmission. }
\section{Precoder Analysis}
\subsection{{Approximate} Spoofed Parameters at Eve}
Herein, we derive {an approximation} on Eve's {parameter estimates} when the DSAS precoder is employed. The approximation is instrumental in predicting Eve's performance {under the assumption that Eve has no knowledge of the DSAS precoder structure}.
We assume that Eve can accurately separate the channel contributions associated with each path, which are denoted by $\widetilde{\bm{H}}^{(n)}_k$ for the $k$-th propagation path. In this case, the channel contribution takes the form
\begin{align}
    \widetilde{\bm{H}}^{(n)}_k 
    =&\gamma_k^\star e^{\frac{-j2\pi n\tau_k^\star}{NT_s}}\sum\limits_{m=0}^{M-1}
    \alpha_{m, k}\bm{A}_{N_R, N_T}(\theta_{Rx, k}^\star, \phi_{Tx, m}^{( n)}).
    \label{eq:spoof_channel_sub_mul}
\end{align}
% where $\alpha_{m, k}\doteq g_{N_T}\left(\frac{2m-M}{M}- \sin\theta_{Tx, k}^\star\right)$.

Then Eve's task is simplified to estimating $\theta_{Rx, k}$ and $\theta_{Tx, k}$ from $ \widetilde{\bm{H}}^{(n)}_k$, $\forall k \in \left\{0, \cdots, K\right\}$. This can be framed as
\begin{subequations}
\begin{align}
\hat{\theta}_{Rx, k}& = & \arg \max\limits_{\theta_{Rx, k}} \left\Vert{\widetilde{\bm{H}}_k^{(n)}\left(\widetilde{\bm{H}}_k^{(n)}\right)^\herm}\bm{a}_{N_R}\left(\theta_{Rx, k}\right)\right\Vert_2^2
\label{eq:aoa_opt}
\\
\hat{\theta}_{Tx, k}& = & \arg \max\limits_{\theta_{Tx, k}} \left\Vert{\left(\widetilde{\bm{H}}_k^{(n)}\right)^\herm\widetilde{\bm{H}}_k^{(n)}}\bm{a}_{N_T}\left(\theta_{Tx, k}\right)\right\Vert_2^2.
\label{eq:aod_opt}
\end{align}
\end{subequations}

By plugging Equations~\eqref{eq:A_mat} and~\eqref{eq:spoof_channel_sub_mul} into the objective function of Equation~\eqref{eq:aoa_opt}, we have,
\begin{align}
    &\left\Vert{\widetilde{\bm{H}}_k^{(n)}\left(\widetilde{\bm{H}}_k^{(n)}\right)^\herm}\bm{a}_{N_R}\left(\theta_{Rx, k}\right)\right\Vert_2\nonumber
    % \\
    % =&\left\Vert\bm{a}_{N_R}\left(\theta_{Rx,k}^\star\right)\left(\bm{b}^{(n)}\right)^\herm\bm{b}^{(n)}\bm{a}^\herm_{N_R}\left(\theta_{Rx,k}^\star\right)\bm{a}_{N_R}\left(\theta_{Rx, k}\right)\right\Vert_2\nonumber
    \\
    =&\left\vert\gamma_k^\star\right\vert N_R\left\Vert\bm{b}^{(n)}\right\Vert^2\left\vert\bm{a}^\herm_{N_R}\left(\theta_{Rx,k}^\star\right)\bm{a}_{N_R}\left(\theta_{Rx, k}\right)\right\vert,
    \label{eq:aoa_est_simplify}
\end{align}
where $\bm{b}^{(n)}\doteq\sum\limits_{m=1}^{M-1}\overline{\alpha}_{m, k}\bm{a}_{N_T}\left(\phi_{Tx, m}^{(n)}\right)$ is not a function of $\theta_{Rx, k}$. From~\eqref{eq:aoa_est_simplify}, we see that the optimal solution to the program~\eqref{eq:aoa_opt} is $\hat{\theta}_{Rx, k}=\theta^\star_{Rx, k}$. 

Similarly, plug Equations~\eqref{eq:A_mat} and~\eqref{eq:spoof_channel_sub_mul} into Equation~\eqref{eq:aod_opt},
\begin{align}
    &\left(N_R\left\vert\gamma_k^\star\right\vert\right)^{-1}{\left(\widetilde{\bm{H}}_k^{(n)}\right)^\herm\widetilde{\bm{H}}_k^{(n)}}\bm{a}_{N_T}\left(\theta_{Tx, k}\right)\nonumber
    \\
    =&\sum\limits_{p=1}^{M-1}\sum\limits_{m=1}^{M-1}\overline{\alpha}_{p, k}\alpha_{m,k}\bm{a}_{N_T}\left(\phi_{Tx, p}^{(n)}\right)\bm{a}_{N_T}^\herm\left(\phi_{Tx, m}^{(n)}\right)\bm{a}_{N_T}\left(\theta_{Tx, k}\right)\nonumber
    % \\
    % =&\sum\limits_{p=1}^{M-1}\overline{\alpha}_{p, k}\bm{a}_{N_T}\left(\phi_{Tx, p}^{(n)}\right)\sum\limits_{m=1}^{M-1}\alpha_{m,k}\bm{a}_{N_T}^\herm\left(\phi_{Tx, m}^{(n)}\right)\bm{a}_{N_T}\left(\theta_{Tx, k}\right)\nonumber
    \\
    =&\left[\sum\limits_{m=0}^{M-1}
     \alpha_{m, k}g_{N_T}\left(\sin\phi^{(n)}_{Tx, k}- \sin\theta_{Tx, k}\right)\right]\bm{b}^{(n)}.
\end{align}

Since $\bm{b}^{(n)}$ is not a function of $\theta_{Tx,k}$, finding the solution to the program~\eqref{eq:aod_opt} is equivalent to solving
\begin{equation}
    \max\limits_{\theta_{Tx, k}} \left\vert\sum\limits_{m=0}^{M-1}
     \alpha_{m, k}g_{N_T}\left(\sin\phi^{(n)}_{Tx, k}- \sin\theta_{Tx, k}\right)\right\vert.
\end{equation}

The magnitude of the function $g_{N_0}\left(\omega\right)$ peaks at $\omega=0$ and quickly diminishes away from $\omega = 0$. Therefore, the AoD estimate can be approximated as $\hat{\theta}_{Tx, k}\approx\phi^{(n)}_{Tx, m_k}$, where $m_k \doteq\argmax\limits_{0 \leq m \leq M-1} \left\vert \alpha_{m,k}\right\vert=  \argmin\limits_{0 \leq m \leq M-1} \min\limits_{t\in \mathbb{Z}}\left\vert 2t+\frac{2m-M}{M}- \sin\theta_{Tx, k}^\star\right\vert$. {This approximation is asymptotically accurate as $N_T\rightarrow \infty$. What is key is that the approximation enables a good spoofing design; even for $N_T=8$ where the approximation is not tight (see Section~\ref{sec:num}), the spoofed AoDs yield high localization error for Eve.
}
% \um{why is this point important?}
% \begin{figure}[!t]
%     \centering
%     \includegraphics[width=0.35\textwidth]{gfunc.eps}
%     \caption{Magnitude of $g_{N_0}\left(\omega\right)$ when $N_0 = 128$.}
%     \label{fig:gfunc}
% \end{figure}

{By properly choosing $\left\{\phi_{k, m_k}^{(n)}\right\}_{k=0}^M$, we can ensure that, almost surely, $\left\vert\phi_{Tx, m_k}^{(n)}-\phi_{Tx, m_0}^{(n)}\right\vert\neq \left\vert\theta^\star_{Tx, k}-\theta^\star_{Tx, 0}\right\vert$, thereby misleading Eve into estimating the wrong ADoDs. The approximation on spoofed ADoDs enables an analysis on the localization error of Eve. Our precoder design is seen to be effective in the numerical results, as motivated by the analysis below.}
% The approximation does not yield a tight lower bound on Eve's performance; however, our precoder design is seen to be effective in the numerical results, as motivated by the analysis below.
% Based on the approximate, we can formulate an approximation of the spoofed channel in Equation~\eqref{eq:spoof_channel_mul} as 
% \begin{align}
%     \widetilde{\bm{H}}^{(n)} \approx &\sum\limits_{k=0}^K
%     \gamma_k^\star \alpha_{m, k}e^{-j2\pi n\frac{\tau_k^\star}{NT_s}}\bm{A}_{N_R, N_T}\left(\theta_{Rx, k}^\star, \phi_{Tx,m_k}^{( n)}\right),
%     \label{eq:spoof_channel_mul_app}
% \end{align}
% which we believe to 
% The approximation retains the components in  $\widetilde{\bm{H}}^{(n)}$ where the arguments of $g_{N_0}(\cdot)$ are closest to zero and thus near maximum.
\vspace*{-0.2in}
\subsection{Bias Calculation for Eve}
% \um{is this something you still have to do?}
For the analysis, we apply the same precoding matrix $\bm{\Psi}\left(\bm{\phi}^{(n)}_{Tx}\right)=\bm{\Psi}\left(\bm{\phi}_{Tx}\right)$ over all subcarriers. Eve will infer $\widetilde{\bm{\tau}} =\bm{\tau}^\star$, $\widetilde{\bm{\theta}}_{Rx}=\bm{\theta}^\star_{Rx}$ and $\widetilde{\bm{\theta}}_{Tx}=\left[\phi_{Tx, m_0}, \cdots, \phi_{Tx, m_K}\right]^\top$ as the parameters of the spoofed channel, where we define $\dot{\phi}_{Tx, m_k}\doteq\left\vert \phi_{Tx, m_k}-\phi_{Tx, m_0}\right\vert$. {While this is an idealized situation for Eve, in practice, the {averaged bias} is close to~\eqref{eq:dist_err}, as shown in our numerical results.}
Denote Eve's estimate of Alice's location 
%While the ground-truth location of Alice $\bm{p}^\star$ is estimated from $\bm{\tau}^\star$, $\bm{\theta}_{Rx}^\star$ and $\bm{\theta}_{Tx}^\star$, Eve will use $\widetilde{\bm{\tau}}$, $\widetilde{\bm{\theta}}_{Rx}$ and $\widetilde{\bm{\theta}}_{Tx}$ to estimate Alice's position 
as $\widetilde{\bm{p}}$. Given the differential localization in Equations~\eqref{eq:ds_loc_original}, \eqref{eq:diff_dist_est_sum} and~\eqref{eq:diff_dist_est_pair}, the error between the spoofed and ground-truth locations {is the error in the distance estimate},
%the spoofed location will differ from the ground-truth location b
\begin{align}
    \left\Vert\bm{p}^\star-\widetilde{\bm{p}}\right\Vert_2 &= \left\vert \sum\limits_{k=1}^K c_k^\star d_k^\star-\sum\limits_{k=1}^K \widetilde{c}_k \widetilde{d}_k\right\vert\overset{(c)}{=} \left\vert \sum\limits_{k=1}^K \widetilde{c}_k d_k^\star-\sum\limits_{k=1}^K \widetilde{c}_k \widetilde{d}_k\right\vert\nonumber
    \\
   & \overset{(d)}{=}\left\vert\sum\limits_{k=1}^K\frac{\widetilde{c}_k{v_c}\dot{\tau}_k^\star}{2}\cot{\frac{\dot{\theta}_{Rx, k}^\star}{2}}(\cot{\frac{\dot{\theta}_{Tx, k}^\star}{2}}-\cot{\frac{\dot{\phi}_{Tx, m_k}}{2}})\right\vert.
    % \\
    % \overset{(b)}{=}&\frac{v_c\dot{\tau}_1}{2}\cot{\frac{\dot{\theta}_{Rx, k}}{2}}\left\vert\frac{\cos{\frac{\dot{\theta}_{Tx, 1}}{2}}}{\sin{\frac{\dot{\theta}_{Tx, 1}}{2}}}-\cot{\frac{\dot{\phi}_{Tx, 1}}{2}}\right\vert
    % \\
    % =&\frac{v_c\dot{\tau}_1}{2}\cot{\frac{\dot{\theta}_{Rx, k}}{2}}\left\vert\frac{\cos{\frac{\dot{\theta}_{Tx, 1}}{2}}\sin{\frac{\dot{\phi}_{Tx, 1}}{2}}-\sin{\frac{\dot{\theta}_{Tx, 1}}{2}}\cos{\frac{\dot{\phi}_{Tx, 1}}{2}}}{\sin{\frac{\dot{\theta}_{Tx, 1}}{2}}\sin{\frac{\dot{\phi}_{Tx, 1}}{2}}}\right\vert
    % \\
    % \overset{(c)}{=}&\frac{v_c\dot{\tau}_1}{2}\frac{\cot{\frac{\dot{\theta}_{Rx, k}}{2}}}{\sin{\frac{\dot{\theta}_{Tx, 1}}{2}}}\left\vert \cos{\frac{\dot{\theta}_{Tx, 1}}{2}}-\sin{\frac{\dot{\theta}_{Tx, 1}}{2}}\cot{\frac{\dot{\phi}_{Tx, 1}}{2}}\right\vert
    % \\
    % \overset{(d)}{\geq}&R_0\left[\sin{\frac{\dot{\theta}_{Tx, 1}}{2}}\left\vert \cot{\frac{\dot{\phi}_{Tx, 1}}{2}}\right\vert-\cos{\frac{\dot{\theta}_{Tx, 1}}{2}}\right],
    \label{eq:dist_err}
\end{align}
where $(c)$ is due to $\sum\limits_{k=1}^K c_k^\star = \sum\limits_{k=1}^K \widetilde{c}_k=1$ and $d_k^\star=d^\star$, for $ k =1, \cdots, K$. Equality $(d)$ follows by plugging in Equation~\eqref{eq:diff_dist_est_pair}.

From Equation $\eqref{eq:dist_err}$, we observe that a non-zero localization bias exists under the spoofed parameters. 
% However, the errors given by different paths might cancel each other when summing up, making it difficult to derive a lower bound on the overall localization error.
% For a special case of $K=1$.
% $(b)$ follows by $\cot{x}=\frac{\cos{x}}{\sin{x}}$. $(c)$ is obtained by factoring out $\sin{\frac{\dot{\theta}_{Tx, 1}}{2}}$ from the absolute value. Inequality $(d)$ holds as $\left\vert x-y\right\vert\geq \left\vert y\right\vert-\left\vert x\right\vert$ and $\frac{\dot{\theta}_{Tx, 1}}{2}\in \left(0, \frac{\pi}{2}\right)$.

We assume $\phi_{Tx, m}$ are randomly sampled from the set $\left\{\sin^{-1}{\left(\frac{2 m+1-M_\phi}{M_{\phi}}\right)}:m=0, \cdots, M_{\phi}-1\right\}$ without replacement;  this ensures that $\bm{\Psi}\left(\bm{\phi}_{Tx}^{(n)}\right)$ is unitary {(and thus full rank)} when $M_\phi=M$. Let $K=1$ in Equation~\eqref{eq:dist_err}. {Note that when $\phi_{Tx,m0}=\phi_{Tx,m1}$, the estimation error is infinity. To develop an interpretable bound, we further assume that $\phi_{Tx,m0}\neq\phi_{Tx,m1}$.}
% , then we can calculate
% \begin{equation}
%     p_{\frac{\dot{\phi}_{Tx, 1}}{2}}\left(\frac{m\pi}{M_{\phi}}\right) = \frac{2(M_{\phi}-m)}{M_{\phi}(M_{\phi}-1)}\text{, for~}m\in \left\{1, \cdots, M_{\phi}-1\right\}.
% \end{equation}
% This leads to
% \begin{equation}
%     p_{\left\vert \cot{\frac{\dot{\phi}_{Tx, m_1}}{2}}\right\vert}\left(\cot{\frac{m\pi}{M_{\phi}}}\right) = \frac{2}{M_{\phi}-1},
%     \label{eq:cot_pmf}
% \end{equation}
% for $m\in \left\{1, \cdots, \left\lfloor \frac{M_{\phi}-1}{2}\right\rfloor\right\}$.
By defining ${R_0}\doteq \frac{v_c\dot{\tau}^\star_1}{2}\cot{\frac{\dot{\theta}^\star_{Rx, 1}}{2}}$,
% and assuming $\frac{m_\theta}{M_\phi}<\frac{\dot{\theta}_{Tx, 1}}{2}<\frac{\left(m_\theta+1\right)}{M_\phi}$ for some $m_\theta\leq \left\lfloor\frac{M_\phi}{2}\right\rfloor$, the lower bound on the expected localization error for Eve writes
{
\begin{align}
    &\mathbb{E}_{{\bm{\phi}_{Tx, m}}}\left[\left\Vert\bm{p}^\star-\widetilde{\bm{p}}\right\Vert_2\right]
    % =R_0\mathbb{E}\left[\left\vert\cot{\frac{\dot{\theta}^\star_{Tx, 1}}{2}}-\cot{\frac{\dot{\phi}_{Tx,m_1}}{2}}\right\vert\right]\nonumber
    % \\
    % =& R_0\sum\limits_{i=1}^3\mathbb{P}\left(E_i\right)\mathbb{E}\left[\left\vert\cot{\frac{\dot{\theta}_{Tx, 1}}{2}}-\cot{\frac{\dot{\phi}_{Tx, 1}}{2}}\right\vert\vert E_i\right],
    \overset{(e)}{\geq}  {R_0}\left[\mathbb{E}\left[ \cot{\frac{\dot{\phi}_{Tx, 1}}{2}}\right]-\cot{\frac{\dot{\theta}^\star_{Tx, 1}}{2}}\right]\nonumber
    \\
    %  \overset{(f)}{\geq} & R_0\left\vert\frac{1}{\left\lfloor \frac{M_{\phi}}{2}\right\rfloor}\left(\sum\limits_{m=1}^{\left\lfloor \frac{M_{\phi}}{2}\right\rfloor}\cot{\frac{m\pi}{M_{\phi}}}\right)-\cot{\frac{\dot{\theta}^\star_{Tx, 1}}{2}}\right\vert \nonumber
    %  \\
    %   \overset{(g)}{>} & R_0\left\vert\frac{1}{\left\lfloor \frac{M_{\phi}}{2}\right\rfloor}\sum\limits_{m=1}^{\left\lfloor \frac{M_{\phi}}{2}\right\rfloor}\left(\frac{M_{\phi}}{m\pi}-\frac{4m}{\pi M_{\phi}}\right)-\cot{\frac{\dot{\theta}^\star_{Tx, 1}}{2}}\right\vert \nonumber
    % \\
     \overset{(f)}{\geq} & {R_0}\left[\left(\frac{M_\phi}{2M_\phi-1} H_{M_{\phi}-1}-2\right)-\cot{\frac{\dot{\theta}^\star_{Tx, 1}}{2}}\right]\nonumber
    \\
     \overset{(g)}{\geq} & {R_0}\left[\Theta\left(\log{M_{\phi}}\right)-\cot{\frac{\dot{\theta}^\star_{Tx, 1}}{2}}\right]  =\Theta\left(\log{M_{\phi}}\right),
     \label{eq:dist_err_exp_lb}
\end{align}
}
% where we define the events $E_1\doteq \left\{\frac{\dot{\phi}_{Tx, 1}}{2}\in \left(0, \frac{m_\theta\pi}{M_\phi}\right]\right\}$, $E_2\doteq \left\{\frac{\dot{\phi}_{Tx, 1}}{2}\in \left(\frac{m_\theta\pi}{M_\phi}, \frac{\pi}{2}\right)\right\}$ and $E_3\doteq \left\{\frac{\dot{\phi}_{Tx, 1}}{2}\in \left[\frac{\pi}{2}, \pi\right)\right\}$.
where $(e)$ follows from Jensen's inequality and $\vert x\vert \geq x, \forall x\in \mathbb{R}$. Inequality $(f)$ holds by the inequality $\cot{\left\vert\frac{\vartheta-\varphi}{2}\right\vert} \geq \frac{2-\left\vert\sin{\vartheta}\right\vert-\left\vert\sin{\varphi}\right\vert}{\left\vert\sin{\vartheta}-\sin{\varphi}\right\vert}$ for $\vartheta,\varphi\in\left(-\frac{\pi}{2}, \frac{\pi}{2}\right)$, and under the assumption of randomly sampling $\sin{\phi_{Tx, m_0}}\neq\sin{\phi_{Tx, m_1}}$ from $\left\{\frac{2 m+1-M_\phi}{M_{\phi}}:m=0, \cdots, M_{\phi}-1\right\}$, where $\mathcal{H}_n\doteq\sum\limits_{i=1}^{n}\frac{1}{i}$. $(g)$ follows from the $n$-th harmonic number $\mathcal{H}_n=\Theta\left(\log{n}\right)$~\cite{guo2011sharp}.
% \um{need a different letter than $H_n$ since you are using $\mathbf{H}$ for your channel.}

To increase the spoofing error at $K=1$, Equation~\eqref{eq:dist_err_exp_lb} suggests that $M_\phi\rightarrow \infty$, which is equivalent to drawing $\sin{\phi_{Tx, k}}\sim \mbox{Uniform}\left(-1, 1\right)$. However, Bob's performance degrades as $M_\phi$ increases in practice. Therefore, a trade-off exists in choosing $M_\phi$: an effective choice is $M_\phi=M=N_T$. {A similar analysis on Eq. $\eqref{eq:dist_err}$ shows that the magnitude of each term in the sum scales with $\Theta\left(\log{M_\Phi}\right)$. }

%{From Equation~\eqref{eq:diff_dist_est_sum}, we figure out the error at $K>1$ can be expressed as the weighted average over multiple errors at $K=1$. Then, we expect the error for $K>1$ to also increase in the rate $\Theta\left(\log{M_\Phi}\right)$.}
% % From Equation~\eqref{eq:dist_err_exp_lb}, we observe that the spoofing error grows unboundedly as $M_{\phi}\rightarrow\infty$.

% As derived in~\eqref{eq:diff_dist_est_sum}, when $K>2$, the distance estimate can be expressed as weighted average over all pairwise distance estimate between the LoS path and NLoS paths. The same lower bound in Equation~\eqref{eq:dist_err_exp_lb} can be adapted to lower-bound the distance bias in $K>2$ case.
\vspace{-10pt}
% \um{so the analysis is missing?}
\section{Simulation Results}
\label{sec:num}
% \um{make clear who knows what, what does Eve know, what does Bob know in your numerical results}
% \um{re do all of the legends, do not use the term "algorithm" but spoofing matrix or precoder.  Find a name for your approach and use the acronym throughout the paper, same with DAIS, DAIS is not an algorithm it is a precoder design}
The following parameters are fixed: $N=81$, $N_R=N_T=8$, $B=30 \mbox{MHz}$. For each channel realization, generate $K+1$ i.i.d. random variables $\tau_k^\star\sim\mbox{Uniform}\left(0, NT_s\right)$, which are relabeled in ascending order as $\tau_0^\star\leq \cdots \leq \tau_K^\star$; generate $\theta_{Tx, k}^\star\sim \mbox{Uniform}\left(-\frac{\pi}{2}, \frac{\pi}{2}\right)$; place Eve/Bob at $[0,0]^\top$, and compute Alice's location and AoAs as $\bm{p}^\star= -v_c\tau^\star_0\left[\cos{\theta_{Tx, 0}}, \sin{\theta_{Tx, 0}}\right]^\top$ and $\theta_{Rx, k}^\star\in\left(-\frac{\pi}{2}, \frac{\pi}{2}\right)$, respectively; generate $\mathrm{\gamma}_k^\star\sim\mathcal{CN}\left(0, \exp{-0.05\frac{\tau_k^\star}{NT_s}}\right)$. For DSAS, set $M=N_T=8$ and $\bm{\phi}_{Tx}^{(n)}=\bm{\phi}_{Tx}$ for $n=1, \cdots, N$, and generate the spoofed AoDs $\bm{\phi}_{Tx}$ from the set $\left\{\sin^{-1}{\left(\frac{2 m+1-M_\phi}{M_{\phi}}\right)}:m=0, \cdots, M_{\phi}-1\right\}$ {without replacement}. 
{As our goal is to evaluate the impact of the spoofing precoder, we exclude channel realizations that yield large localization errors for the {\bf unspoofed} case.}

%{Under the above channel generation strategy, some channel realizations will exhibit large localization errors even in the absence of spoofing against Eve. To better evaluate the spoofing performance of our method, we exclude these rare cases, which account for approximately $16\%$ of the total.} 

For DAIS, we adapt the precoding matrix $\bm{\Phi}^{(n)}\left(\Delta_\tau=0.5NT_s, \Delta_\theta=0.25\pi\right)$ according to~\eqref{eq:dais_mat}. {For DAIS\_a, we set the precoding matrix as $\bm{\Phi}^{(n)}\left(0, 0.25\pi\right)$, with only angle spoofing.} For DAIS+DSAS, we use the product of the precoders as  $\bm{\Phi}^{(n)}\left(0.5NT_s, 0.25\pi\right)\bm{\Psi}\left(\bm{\phi}_{Tx}\right)$ on the $n$-th subcarrier. We compare the performances of above spoofing strategies under the non-differential (ND) localization method used in~\cite{li2025delay}.

The evaluation metric is 
% the normalized root mean square error normalized by the ground-truth distance between the transmitter and receiver
: $\mbox{NEE}=\mathbb{E}\left[\frac{\left\Vert\bm{p}^\star-\hat{\bm{p}}\right\Vert_2}{\left\Vert \bm{p}^\star-\bm{q}^\star\right\Vert_2}\right]$, where the expectation is taken w.r.t. channel, precoder and noise realizations. {The precoder matrix $\bm{\Psi}\left(\bm{\phi}_{Tx}\right)$ is updated for each transmission.} {Let $\bm{W}\in \mathbb{C}^{N_R\times N_R}$ and $\bm{F}\in \mathbb{C}^{N_T\times N_T}$ be unitary matrices, and assume all symbols have unit energy, i.e. $\left\vert S^{(n)}_{r_1, r_2}\right\vert=1$}. The signal-to-noise ratio is defined as $\mbox{SNR(dB)} =10\log_{10}\left[\sum\limits_{n=1}^N\left\Vert\bm{H}^{(n)}\right\Vert_F^2/\left(NN_RN_T\sigma^2\right)\right]$,
% $\mbox{SNR(dB)} =10\log_{10}{\frac{\sum\limits_{n=1}^N\left\Vert\bm{H}^{(n)}\right\Vert_F^2}{NN_RN_T\sigma^2}}$,
% \begin{align}
%     \mbox{SNR(dB)} &=10\log_{10}{\frac{\sum\limits_{n=1}^N\left\Vert\bm{H}^{(n)}\right\Vert_F^2}{NN_RN_T\sigma^2}},
%     % \\
%     % \mbox{Spoofing: SNR(dB)} &=10\log_{10}{\frac{\sum\limits_{n=1}^N\left\Vert\widetilde{\bm{H}}^{(n)}\right\Vert_F^2}{NN_RN_T\sigma^2}}.
% \end{align}
where $\bm{H}^{(n)}$ is replaced by $\widetilde{\bm{H}}^{(n)}$ for spoofed channels. {We estimate the parameters using FLEX~\cite{pourafzal2025flex} with correlation-based angle estimation. We assume that the number of paths is unknown and all the paths have distinct AoAs and AoDs.}
\vspace{-10pt}
\subsection{Impact of $\mbox{SNR(dB)}$}
Figure~\ref{fig:DAIS_exp_snr} shows that Bob's performance overlaps with that under non-spoofing channels, verifying that the spoofing effect can be eliminated with the side information. The bias introduced by DAIS and DSAS dominates Eve's errors, causing Eve's curves to appear nearly flat. As expected, {DAIS\_a} introduces localization errors greater than {$0.5$} against Eve in only around {61\%} of 236 channel realizations, whereas the normalized localization errors produced by DSAS and DAIS+DSAS exceed {$0.5$ for around $99\%$} channel realizations. 
% However, the average performance of DAIS is superior to that of DSAS.
By combining the two strategies, DAIS+DSAS achieves the best overall performance, increasing Eve's localization error by approximately {23$\%$} compared with DAIS. Under the ND localization method (dashed lines), DSAS and DAIS+DSAS schemes are still effective.
\begin{figure}[!t]
    \centering
    \vspace*{-0.1in}
    \includegraphics[width=0.42\textwidth]{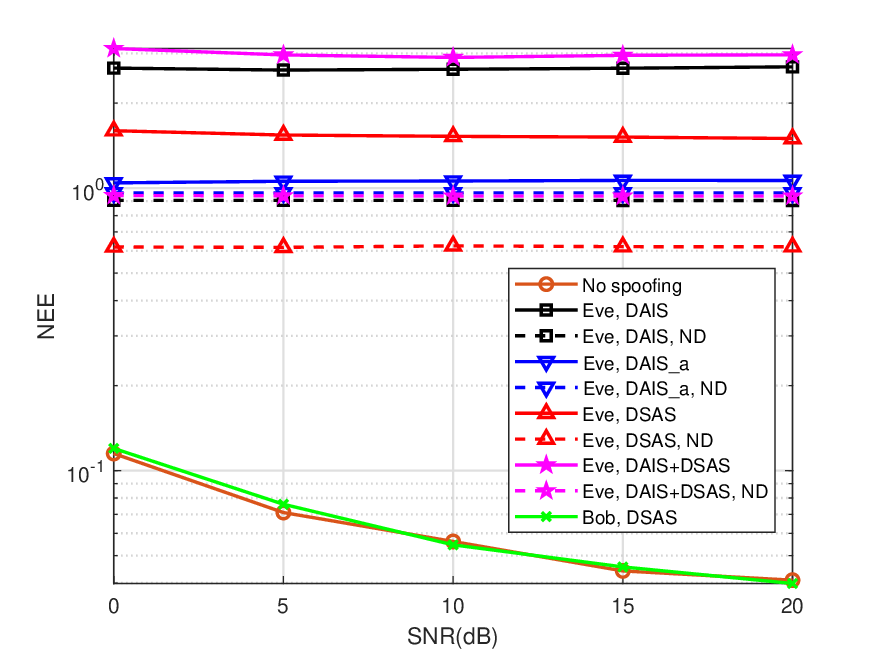}
    \vspace{-10pt}
    \caption{NEE results versus $\mbox{SNR(dB)}$. Set $M_{\phi}=8$, $K=2$.}
    \label{fig:DAIS_exp_snr}
    \vspace{-12pt}
\end{figure}
\vspace{-10pt}
\subsection{Impact of $M_\phi$}
In Figure~\ref{fig:DAIS_exp_mphi}, we investigate into the impact of the minimum separation between the sine of spoofing AoDs, $\frac{2}{M_\phi}$, on the spoofing performance. The error predicted by Equation~\eqref{eq:dist_err} approaches Eve's performance under DSAS. As indicated by~\eqref{eq:dist_err_exp_lb}, the averaged error introduced by DSAS increases slowly with $M_\phi$. However, as $M_\phi$ increases, Bob's performance degrades rapidly, because the condition number of $\bm{\Psi}\left(\bm{\phi}_{Tx}^{(n)}\right)$ increases, resulting in noise amplification when Bob reverses the spoofing effect. It suggests that we should choose $M_\phi=M=N_T$, where $\bm{\Psi}\left(\bm{\phi}_{Tx}^{(n)}\right)$ becomes a unitary matrix.
\begin{figure}[!t]
    \centering
    \includegraphics[width=0.42\textwidth]{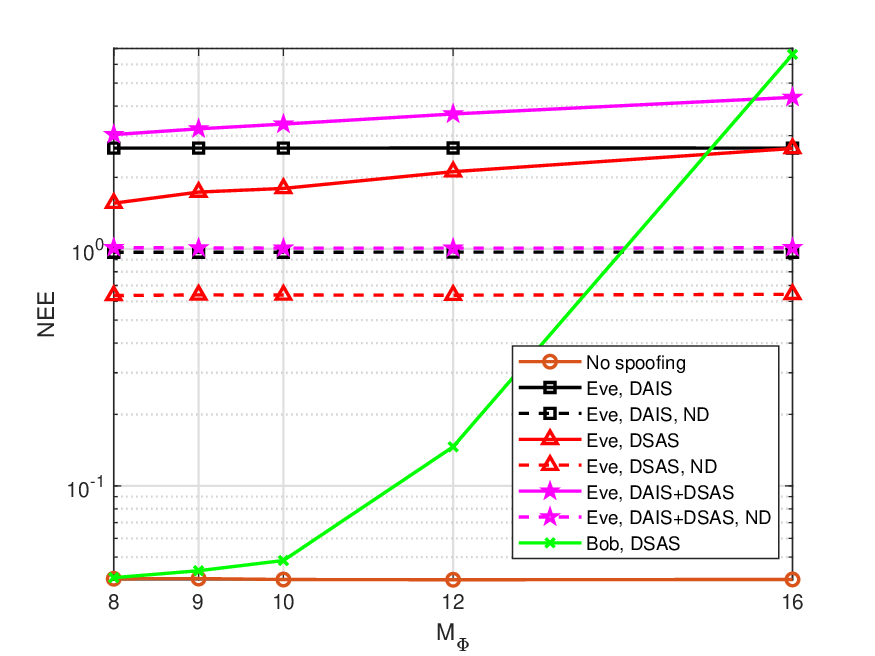}
    \vspace{-10pt}
    \caption{NEE results versus $M_\phi$. Set $\mbox{SNR}=20\mbox{dB}$, $K=2$.}
    \label{fig:DAIS_exp_mphi}
    \vspace{-17pt}
\end{figure}
% \subsection{Impact of the number of NLoS paths, $K$}
% As shown in Figure~\ref{fig:DAIS_exp_k}, 
% \begin{figure}[!t]
%     \centering
%     \includegraphics[width=0.42\textwidth]{exp_l_final.eps}
%     \caption{NRMSE results versus the number of NLoS paths, $K$. Set $M_{\phi}=8$, $\mbox{SNR}=20\mbox{dB}$.}
%     \label{fig:DAIS_exp_k}
% \end{figure}
% \um{For Figure~\ref{fig:DAIS_exp} make clear what happens for $K=1,2$}
\vspace{-17pt}
\section{Conclusions}
In this work, we proposed a CSI-free precoding matrix design, which distorts Eve's channel estimation, while allowing Bob to reverse the spoofing effect. {By applying the DSAS precoder}, we analyzed the channel parameters that Eve infers from the spoofed channel, demonstrating that {DSAS} produces spoofed ADoDs {for Eve}. This confirms the effectiveness of {DSAS} under the differential-signaling localization, where DAIS will {be less effective} unless phase wrapping occurs. With the {approximate} spoofed parameters, we derived the localization bias introduced into Eve's location estimate. Simulation results verify that the DSAS strategy introduces a significant localization bias into Eve's estimates. {When combined with DAIS, DSAS achieves the best overall spoofing performance.}
{
\vspace{-20pt}
\section*{Acknowledgments}
We wish to thank Prof. H. Wymeersch for thoughtful discussions on differential localization that motivated this work.}
\renewcommand*{\bibfont}{\normalfont\footnotesize}
\newpage
\printbibliography
% \section{TODO}
% \begin{enumerate}
% \item Change the legends of your plots, we are not defining new algorithms, but new precoders
% \item might need to explain the differential localization approach of [1] a bit
% \item put in full proof for $K = 2$
% \item $K >2$ derivation/discussion
% \item simulation results: add in comparison methods, performance versus minimum separation, SNR -- also add in numerical results for $K > 2$
% \end{enumerate}
\end{document}